\documentclass[conference]{IEEEtran}
\IEEEoverridecommandlockouts
\usepackage{amsmath,amssymb,amsthm}
\usepackage{booktabs}
\usepackage{algorithm}
\usepackage[noend]{algpseudocode}
\usepackage{graphicx}
\usepackage{cite}
\usepackage{xurl}
\usepackage[hidelinks]{hyperref}

\newtheorem{proposition}{Proposition}

\theoremstyle{remark}

\newcommand{\ssd}{\mathrm{ssd}}
\newcommand{\pf}{\mathrm{pf}}
\newcommand{\nrate}{\rho_{\mathrm{rec}}}

\usepackage{xcolor}
\begin{document}

\title{AstraLLM: Serving Multi-Turn LLM Sessions\\Across Orbital Data-Center Satellites}
\author{\IEEEauthorblockN{Hung-chun Lin\IEEEauthorrefmark{1}, Ting-Wei Hsu\IEEEauthorrefmark{1}, Ching-Lun Tai\IEEEauthorrefmark{2}, Karthikeyan Sundaresan\IEEEauthorrefmark{1}, Raghupathy Sivakumar\IEEEauthorrefmark{1}}
\IEEEauthorblockA{\IEEEauthorrefmark{1}Georgia Institute of Technology \qquad \IEEEauthorrefmark{2}Independent Researcher}}
\maketitle

\begin{abstract}
As the power demand of large language models (LLMs) surges, data centers are moving into space to harness abundant solar power.
However, serving multi-turn LLM conversations in orbit breaks a foundational assumption of terrestrial serving systems, which rely on stationary servers to reuse each session's key-value (KV) cache.
In space, the satellite holding a session's KV cache continuously drifts away between user turns, becoming a \textit{drifting holder}.
Although this drift adds only minor latency to a single turn, naive scheduling decisions severely overload scarce onboard GPUs and congest inter-satellite links, degrading performance for later requests.
We present AstraLLM, the first scheduler designed for multi-turn LLM serving across orbital data centers.
AstraLLM jointly selects the serving satellite and KV-cache action for each turn by pricing the congestion its GPU work and cache transfers impose on subsequent requests, while probabilistically spreading traffic across independent gateways.
In a packet-level ns-3 simulation built from real Starlink orbits and production LLM traces, AstraLLM reduces mean time to first token (TTFT) by 2.5$\times$ compared with the strongest serving system, and serves 1.7$\times$ and 2.3$\times$ more requests per second than SGLang and Mooncake.
\end{abstract}

\section{Introduction}
\label{sec:intro}

\begin{figure}[t]
\centering
\includegraphics[width=\columnwidth]{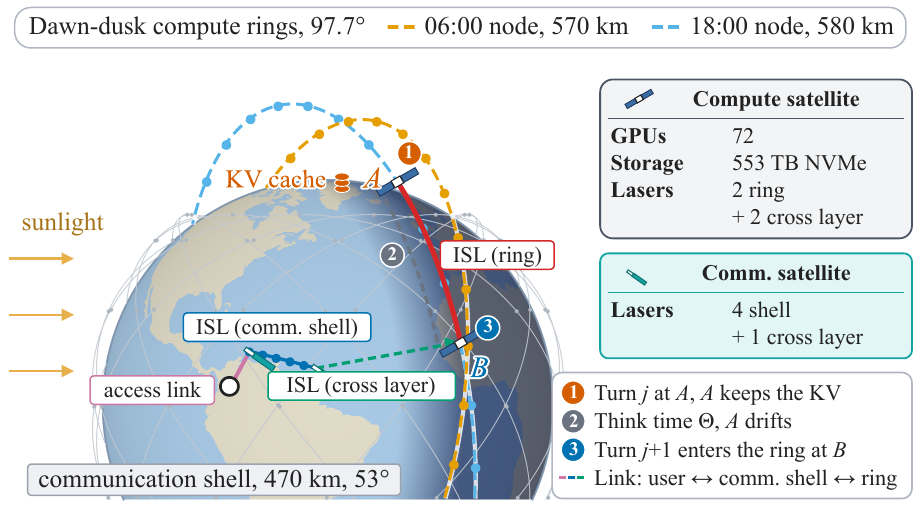}
\caption{System overview.  Two dawn-dusk
compute rings fly above the $53^\circ$ communication shell of
Starlink, whose relays carry every turn to the rings.  Turn $j$ of a session from Bogot\'a
was served at $A$, which keeps its KV cache (1).  During the think
time the ring carries $A$ away (2), so turn $j+1$ enters the ring at
$B$, several ring hops from $A$ (3).}
\label{fig:overview}
\end{figure}

Global data-center electricity demand is projected to more than double---from 415~TWh in 2024 to roughly 945~TWh in 2030~\cite{iea2025energyai}---driving more than one-fifth of all new power demand in advanced economies, largely fueled by large language model (LLM) workloads~\cite{msft2025aidiffusion}. To overcome terrestrial power bottlenecks, technology companies are moving data centers into space to harness abundant solar power. Satellites in a dawn-dusk Sun-synchronous orbit (SSO), which follows the boundary between day and night, receive near-continuous sunlight and uninterrupted solar power~\cite{aguera2026,spacex2026grant}. This vision is moving from study to deployment. Starcloud flew an NVIDIA H100 GPU aboard Starcloud-1 in November 2025~\cite{starcloud2025h100,nvidia2025starcloud}, and Google launched its first Project Suncatcher prototype carrying tensor processing units (TPUs) in October 2026~\cite{aguera2026,beals2026suncatcher}. In November 2026, the Federal Communications Commission (FCC) further authorized SpaceX to deploy an orbital data-center system of up to one million satellites, with milestones for the first satellite by 2033 and 10\% of the constellation by 2035~\cite{spacex2026grant}. Because hosting GPU racks requires dedicated power, thermal, and structural designs that existing communication satellites like Starlink cannot support, operators must launch dedicated compute satellites into separate rings above existing communication shells (Fig.~\ref{fig:overview}). Among the workloads these orbital data centers aim to support, one is particularly important given its rapidly growing computational demand: LLM inference.

Modern LLM serving systems commonly adopt prefill-decode (PD) disaggregation, separating prompt processing (prefill) from token-by-token generation (decode) to maximize GPU efficiency~\cite{zhong2024distserve}. Prefill computes Transformer's attention over the entire context to build the key-value (KV) cache, dominating the time to first token (TTFT). Because most LLM traffic consists of multi-turn conversations that prepend prior turns to each new prompt~\cite{wang2025kvwild}, keeping the KV cache lets a follow-up turn skip redundant prefill. Terrestrial serving systems are therefore designed around cache reuse. SGLang routes to the longest cached prefix unless loads become unbalanced~\cite{zheng2024sglang}, while Mooncake transfers caches across instances to minimize estimated TTFT~\cite{qin2025mooncake}. All of these systems assume stationary terrestrial servers with uniform distance to the user. Without a distance penalty to reach the cache, reuse is therefore both faster for the turn and cheaper for the system.

Orbital data centers break this stationary assumption because their servers move continuously in space. On a dawn-dusk ring of 288 satellites at 570~km altitude, each satellite travels at 7.6~km/s. During a user's median think time of 106~s~\cite{wang2025kvwild}, the satellite that served the previous turn—the holder—drifts 800 km along the ring, roughly five satellites away, as Fig.~\ref{fig:overview} illustrates. We call this phenomenon the \textit{drifting holder}. When the next turn arrives, the holder is no longer the satellite nearest to the user (the entry satellite). The gateway must therefore choose to either forward the turn to the holder (return), transfer the KV cache over the ring lasers to the entry satellite (fetch) or recompute the full context there (recomputation). We refer to return and fetch together as reuse.

One would expect the drift to make reuse slow. Over the ring lasers, however, reaching the holder or fetching its cache adds only a few milliseconds. This delay is about the prefill time that reuse saves for medium context length, so the three actions often give a similar TTFT per turn (Section~\ref{sec:actions}). Although the three actions look similar in a single turn, they burden the constellation very differently. First, a router that picks each turn's fastest option ignores the wait that its GPU work adds for the requests queued behind it. Each satellite carries only 72 GPUs~\cite{starmind2026nvidia}, causing queues to build up quickly. Second, once GPU work is priced, fetching a cache often looks cheaper than recomputing it or waiting at a busy holder. However, each fetch sends hundreds of megabytes over the ring inter-satellite links (ISLs) and delays the other transfers that also share the same link. Third, because load reports are updated periodically and quickly go stale, a satellite that appears lightly loaded between updates often draws simultaneous requests from multiple gateways. These turns then pile up in a single GPU queue, especially under high load.

To address this load problem, we propose \textbf{AstraLLM}, the first scheduler for multi-turn LLM serving in orbital data centers.
For each turn, AstraLLM jointly selects the serving satellite and its KV-cache action (return, fetch, or recomputation), allowing cache locality to naturally guide placement alongside system load.
Each satellite--action pair is scored by the turn's own TTFT plus two parameter-free congestion prices.
The GPU price charges the extra wait that the turn's prefill imposes on later requests at that satellite, while the link price charges the delay that a fetch adds to other transfers on the ring ISLs.
AstraLLM then probabilistically selects an option, favoring lower scores.
The degree of randomization scales with report age, spreading traffic across satellites so that gateways acting on the same stale reports do not crowd a single one.

This paper makes three contributions.
\begin{itemize}
    \item We identify the drifting holder in orbital LLM serving and show that it is a problem of load rather than latency. Because the three actions yield comparable latency but vastly different GPU and link loads, latency-only routing lets queues build up on the busiest satellites.
    \item We design AstraLLM, a joint satellite-and-action scheduler that scores each option by the turn's TTFT plus parameter-free prices for the delays it imposes on satellite GPUs and ring ISLs. It further randomizes selection in proportion to report staleness so that gateways do not crowd the same satellite.
    \item We evaluate AstraLLM in a packet-level ns-3 simulator serving production multi-turn traces from 260 Starlink gateways through the real $53^\circ$ relay shell, modeling Kimi K2 prefill on satellite GPUs. AstraLLM reduces mean TTFT by 2.5$\times$, 2.9$\times$, and 4.7$\times$ on average compared with SGLang, Mooncake, and llm-d, respectively. It also serves 1.7$\times$ and 2.3$\times$ the request rate of SGLang and Mooncake, cutting the constellation size required for a target workload by at least a third.
\end{itemize}
\section{System Model}
\label{sec:model}

\subsection{Two Compute Rings}
\label{sec:rings}

We organize the orbital compute layer into two distributed dawn-dusk rings~\cite{spacex2026grant} (Table~\ref{tab:params}). Because only two dawn-dusk Sun-synchronous orbital planes exist (crossing the equator northbound at 06:00 and 18:00 local time), populating both maximizes near-continuous solar power. Within each ring, satellites are spaced evenly rather than packed into a tight formation like Google's 81-satellite Suncatcher~\cite{aguera2026}. Tight formations are necessary for distributed training to exchange massive weights and gradients across satellites. In LLM inference, however, the heaviest KV cache transfers stay within a single satellite (Section~\ref{sec:gpu}), so cross-satellite migration does not require tight clustering. Following FCC filings~\cite{spacex2026grant}, we therefore distribute satellites evenly along each ring to sustain continuous global coverage. Because the two rings travel in opposite directions, satellites from different rings pass each other too quickly to establish laser links. A session's KV cache can therefore migrate only along its holder's ring. Within a ring, the lasers operate at $C_e = 400$~Gb/s~\cite{starlink2026v3} with a per-hop delay of $\tau_h$.

\subsection{Reaching the Rings}
\label{sec:access}
Because compute satellites fly only along the day--night line and at higher altitudes above the communication shell, direct ground access to the compute rings is impractical. Following SpaceX's FCC filing~\cite{spacex2026grant} and Starcloud~\cite{carroll2026starcloud}, user requests therefore reach the rings through Starlink's global relay shell via laser links. From a gateway $u$, each turn uplinks to a satellite in Starlink's $53^\circ$ shell (which we call a \emph{relay}), traverses the inter-relay laser mesh, and enters the compute rings over a cross-layer laser~\cite{guo2025enabling}. Each compute satellite carries two cross-layer lasers and each relay carries one (Fig.~\ref{fig:overview}). Following Hao et~al.~\cite{hao2025interlayer}, every cross-layer laser connects to the nearest available relay and hands off only when a closer relay appears. Because a turn is only a few kilobytes, forwarding it consumes negligible laser bandwidth.
The shortest one-way latency from gateway $u$ to compute satellite $v$ is therefore
\begin{equation}
\ell(u, v) = \min_{v'} \bigl[ \ell^{\mathrm{cl}}(u, v') + h(v', v)\, \tau_h \bigr],
\label{eq:fwd}
\end{equation}
where $v'$ ranges over the satellites in $v$'s ring, $\ell^{\mathrm{cl}}(u, v')$ is the propagation delay from gateway $u$ through the relays into $v'$ over its own cross-layer laser, and $h(v', v)$ is the number of ring hops from $v'$ to $v$. We call the compute satellite with the minimum latency the gateway's \emph{entry satellite}, and both the entry satellite and its latency change continuously as the satellites move.

\subsection{Prefill Work and Load}
\label{sec:gpu}

Following recent orbital data-center designs~\cite{starmind2026nvidia}, each compute satellite houses a 72-GPU NVLink rack, allowing it to apply PD disaggregation~\cite{zhong2024distserve} internally without transferring KV caches over slower ISLs. Its $J$ prefill groups are tensor-parallel GPU groups that each host a model replica and batch waiting requests via chunked prefill. The remaining GPUs form one decode instance with data-parallel attention and expert parallelism~\cite{deepseek2025infer}, and each decode step takes the time to read the weights and caches from high-bandwidth memory (HBM) (Section~\ref{sec:setup}). Since decoding cost is unaffected by routing or cache reuse, prefill is the only work that varies across scheduling policies and is measured in seconds of a prefill group.

Reusing the prior turns' KV cache (prefix caching) allows a satellite to skip the conversation history and prefill only the new tokens. Prefilling $m$ new tokens after $n$ cached tokens takes
\begin{equation}
  T_{\pf}(n, m) = b\, m + c\, m \bigl(n + m/2\bigr) ,
  \label{eq:tpf}
\end{equation}
where $b$ is the model weight computation time per token in Transformer, and $c$ is the attention time per token pair, both derived from the operation counts of the model (Table~\ref{tab:params}, Section~\ref{sec:setup}). Because the attention term grows quadratically with the number of processed tokens, recomputing a long conversation from scratch is much slower than prefilling only the new turn.

We measure system utilization using two load metrics, the per-group offered load $\rho_g$ and the constellation-wide all-recompute load $\nrate$. The offered load $\rho_g$ is the prefill work arriving at a group per second, so the group's queue grows without bound whenever $\rho_g > 1$. At the constellation level, the all-recompute load $\nrate$ is the ratio of total prefill work to aggregate prefill capacity, assuming every turn recomputes all history and no KV cache is reused. Because the all-recompute load ignores cache hits, $\nrate$ depends only on the request rate and provides a policy-independent measure of demand. Once $\nrate$ exceeds one, the constellation can sustain incoming demand only by reusing KV caches.

\subsection{Return, Fetch or Recompute}
\label{sec:actions}
\begin{figure}[t]
\centering
\includegraphics[width=\columnwidth]{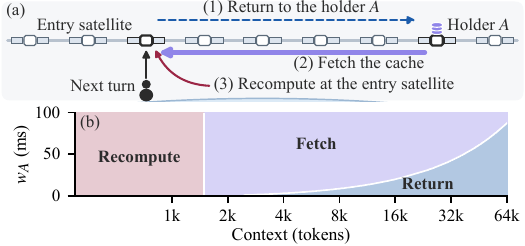}
\caption{(a)~The three actions.  (b)~The fastest action for Kimi K2 on Rubin GPUs, with the holder 800~km away over idle ring lasers and an idle entry satellite.}
\label{fig:fastest}
\end{figure}
\begin{figure*}[t]
\centering
\begin{minipage}[b]{1.70in}\centering\includegraphics{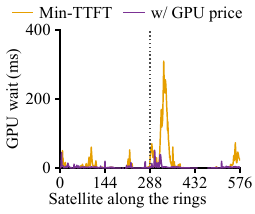}\\[-4pt]{\footnotesize (a)~GPU groups, $\nrate = 0.49$}\end{minipage}\hfill
\begin{minipage}[b]{3.24in}\centering\includegraphics{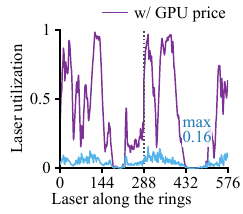}\includegraphics{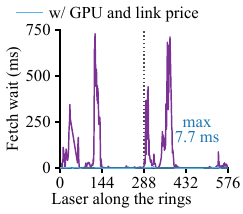}\\[-4pt]{\footnotesize (b)~Ring lasers, $\nrate = 0.49$}\end{minipage}\hfill
\begin{minipage}[b]{1.96in}\centering\includegraphics{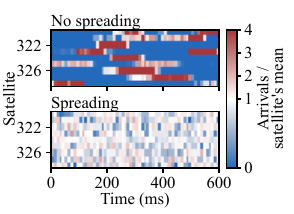}\\[-4pt]{\footnotesize (c)~Gateways acting together, $\nrate = 1.22$}\end{minipage}
\caption{Three scheduling problems in orbit.  (a)~GPU wait per satellite, (b)~utilization of each ring laser and wait of the fetches bottlenecked there, (c)~arrivals per 10~ms over each satellite's mean.}
\label{fig:problems}
\end{figure*}
The holder satellite $A$ stores the KV cache of a session's last turn on its local drives, where its prefill groups can read the cache at rate $R_\ssd$. Because the system prompts shared across many sessions are replicated on every satellite, the first $n_0$ tokens of any turn are always cached and even a full recomputation begins after them.

When the next turn arrives with new tokens from the previous answer and the user's latest message, it can be served at any compute satellite through a return, fetch, or recompute action (Fig.~\ref{fig:fastest}a). At a candidate satellite $v$, an arriving request is assigned to whichever prefill group frees up first, so its queueing delay $w_v$ is the minimum wait across the satellite's groups. The time to first token (TTFT) for executing action $x$ at satellite $v$ is then
\begin{equation}
  \begin{aligned}
  &T^{x}(v) = 2\ell(u, v) \\
  &\; + \begin{cases}
    \max\{ w_v, S/R_\ssd \} + T_{\pf}(n, m), & x = \mathrm{return}, \\
    \max\{ w_v, D_\pi(S) \} + T_{\pf}(n, m), & x = \mathrm{fetch}, \\
    w_v + T_{\pf}(n_0, n{+}m{-}n_0), & x = \mathrm{recompute}.
  \end{cases}
  \end{aligned}
  \label{eq:actions}
\end{equation}

\textbf{Return.} The turn executes at the holder $A$, which reads the $S$-byte KV cache from its local drives and overlaps this read with the queueing wait $w_v$.

\textbf{Fetch.} Once the request arrives, another satellite on the holder's ring serves the turn by pulling the KV cache from $A$ over the shorter ring path $\pi$ in time $D_\pi(S)$. 

\textbf{Recompute.} Without the session's KV cache, the serving satellite reuses only the $n_0$ system-prompt tokens and must prefill all $n{+}m{-}n_0$ remaining tokens after waiting $w_v$, incurring a much longer prefill time $T_{\pf}(n_0, n{+}m{-}n_0)$.

Consequently, the holder serves the turn via return, other satellites on the same ring can fetch or recompute, and satellites on the opposite ring can only recompute. Once the turn completes, the serving satellite becomes the holder for the next turn.

Each gateway schedules its own turns independently without an orbital controller or inter-gateway coordination, as a centralized controller would sit far from most gateways and add prohibitive round-trip delay to every turn. Instead, gateways observe constellation state through periodic load reports that each compute satellite broadcasts every $T_{\mathrm{rep}}$ seconds, summarizing the load of its prefill groups and incoming ring lasers. Because each report arrives after a one-way propagation delay, gateways always schedule turns using delayed constellation state.

As shown in Fig.~\ref{fig:fastest}b, each action achieves the lowest TTFT in a distinct region determined by the prior context length and the holder's queueing wait $w_A$. Recomputation wins for short contexts before growing too costly at longer lengths, where returning is fastest under low $w_A$ and fetching takes over as waiting at the holder increases.

\section{Three Problems of Load}
\label{sec:problems}

Serving LLM inference from a constellation poses three scheduling problems that a ground cluster does not face.  On the ground, a router can send each request to the server with the lowest predicted TTFT~\cite{qin2025mooncake}.  It works there because servers sit about equally far from the users, and the router has the servers information in real time.  Fig.~\ref{fig:problems} shows where this rule fails in orbit, with panels (a) and (b).

\textbf{GPU groups.}  In a ground data center, servers have uniform network distance to users, so reusing a KV cache both reduces latency for the current request and uses less GPU compute. In orbit, however, reaching a distant holder adds transmission delay, so a greedy Min-TTFT rule often favors recomputing on a closer satellite. This choice trades transmission time for heavy GPU computation, overloading a satellite's few prefill groups and forcing subsequent requests to wait.  As Fig.~\ref{fig:problems}a shows, Min-TTFT causes sharp GPU wait spikes on a small subset of satellites, resulting in an average wait of 82~ms.  When the scheduler also accounts for the wait that each request's GPU work adds to later arrivals (Section~IV-A), the average wait drops to 3.5~ms.

\textbf{Ring lasers.}  When GPU load on a satellite is high, the satellite can fetch the KV cache from the holder, shifting the load from GPUs to the ring ISLs.  Unlike ground data-center networks such as Clos or multi-dimensional torus topologies that provide many parallel paths, orbital caches can travel in only two directions along the holder's ring, meaning each fetch effectively has only one path along the shorter arc. Therefore, estimating a fetch's latency only from the KV cache size and hop count will ignore other transfers on the same links, which can easily cause congestion. As Fig.~\ref{fig:problems}b shows, this naive estimation leaves the worst laser nearly full, causing fetches bottlenecked there to wait 730~ms on average.  Accounting for the delay that each fetch adds to other transfers (Section~IV-B) keeps this average wait at most 7.7~ms on every laser.

\textbf{Gateways acting together.}  In a ground data center, a single centralized scheduler can route every request with an up-to-date view of every queue.  In a satellite network, however, distributed gateways cannot observe each other's choices in time and must route independently using periodic, delayed load reports \cite{guo2025enabling}.  Between two reports, all gateways read the same stale numbers and send their turns to whichever satellite looks cheapest.  Even when both GPU and laser delays are accounted for, a satellite can receive up to 9.4 times its mean arrivals for about one report period and none after, leaving 59\% of the plotted time slots empty (Fig.~\ref{fig:problems}c). Randomizing gateway selection can spread traffic and reduce load spikes on individual satellites (Section~IV-C).

To address these three problems together, we propose \textbf{AstraLLM}, a scheduler that jointly accounts for GPU queuing, ring ISL contention, and randomized gateway selection.

\section{AstraLLM}
\label{sec:design}
\label{sec:score}

For each turn, AstraLLM considers every option $(v, x)$ of a compute satellite $v$ and an available action $x$, assigning each option the score
\begin{equation}
  s(v, x) = T^{x}(v) + E^{J}_v\bigl(y(x)\bigr) + E^{\mathrm{link}}_\pi(S)\, \mathbf{1}[x = \mathrm{fetch}] ,
  \label{eq:decision}
\end{equation}
where $T^{x}(v)$ is the TTFT of the turn under the option, from~\eqref{eq:actions}.
The GPU price $E^{J}_v(y(x))$ is the extra wait that the prefill work $y(x)$ of the action adds for later requests at satellite $v$ (Section~\ref{sec:gpuprice}).
The link price $E^{\mathrm{link}}_\pi(S)$ counts only for a fetch and is the extra wait that moving the $S$ bytes of the KV cache along the ring path $\pi$ adds for other transfers (Section~\ref{sec:linkprice}).
Since all gateways share the same state information of the satellites, choosing the lowest score at every gateway would send their requests to the same satellite at once and overload that satellite.
Moreover, the load reports are already stale when a turn arrives, and the GPU price in each score is therefore an estimate with an error.
AstraLLM therefore draws the serving option at random with probability
\begin{equation}
  \Pr\{(v, x)\} \propto e^{-s(v, x)/\beta}, \qquad \beta = \frac{\sqrt{6}}{\pi}\,\theta ,
  \label{eq:draw}
\end{equation}
where $\theta$ is the standard deviation of the error that stale load reports cause in the estimated GPU price (Section~\ref{sec:spread}), and $\beta$ is the corresponding Gumbel scale parameter.
Options whose scores differ by less than about $\beta$ are thus treated as nearly equal and are the candidate options for gateways to choose from.
Both prices charge each choice the extra delay that the choice causes for later requests, known as the marginal congestion cost.
We detail the derivation in Sections~\ref{sec:gpuprice} and~\ref{sec:linkprice}.

Both the TTFT and the GPU price are computed from $w_v$, the time that a turn waits at satellite $v$, but the load reports that the gateway receives from the satellites are always stale.
The latest satellite report is $\alpha_v$ old when the turn arrives, and AstraLLM estimates the wait of each prefill group $g$ of satellite $v$ in three ways.
\textbf{1)~Projected unfinished work.}
The latest report gives the unfinished prefill work $w^{\mathrm{rep}}_g$ of the group.
During the $\alpha_v$ seconds since the report, $\rho_g \alpha_v$ seconds of work arrived at load $\rho_g$ and the group finished $\alpha_v$ seconds of work.
The unfinished work of the group when the turn arrives is therefore $w^{\mathrm{rep}}_g + (\rho_g - 1)\,\alpha_v$.
\textbf{2)~Steady-state wait.}
The Pollaczek--Khinchine formula estimates the long-run mean wait of the group, $q_g / (2(1 - \rho_g))$, rather than relying on a single report as the projected unfinished work does, where $q_g$ is the arrival rate of requests times the mean squared work of a request.
\textbf{3)~Unreported sends.}
The latest report does not yet include the work that the gateway itself has sent to satellite $v$.
Let $W^{u}_v$ be the part of this work that is still unserved when the turn arrives, counting its service from the moment it reaches the satellite.
Spread evenly over the $J$ groups, this work keeps each group busy for at least $W^{u}_v / J$.
For each group, the gateway uses the largest of the three estimates as the estimated waiting time.
When a new request joins satellite $v$, the request goes to the group that frees first, and the gateway therefore estimates the wait at satellite $v$ as
\begin{equation}
  w_v = \min_{g} \max\Bigl\{ w^{\mathrm{rep}}_g + (\rho_g - 1)\,\alpha_v ,\
  \frac{q_g}{2\,(1 - \rho_g)} ,\ \frac{W^{u}_v}{J} \Bigr\} .
  \label{eq:wait}
\end{equation}

\subsection{A GPU Price for the Prefill Groups}
\label{sec:gpuprice}

We define the GPU price as the extra wait that the prefill work of a turn causes for later requests in the same GPU group.
Modeling each group as a first-come-first-served queue with Poisson arrivals, we compute this extra wait in closed form~\cite{haviv1998externalities}, as stated in Proposition~\ref{prop:price}.

\begin{proposition}[Delay imposed by admitted work]
  Admitting $y$ seconds of work to a group with $w$ seconds of
  unfinished work raises the expected total wait of the later arrivals
  by $E_g(y)$, where
  \begin{equation}
    E_g(y) = \frac{\lambda_g\, y\, (w + y/2)}{1 - \rho_g} ,
    \label{eq:ext}
  \end{equation}
  where $\lambda_g$ is the arrival rate of requests at group $g$ and $\rho_g < 1$ is the load of group $g$.
  \label{prop:price}
\end{proposition}

\begin{proof}
  Consider the queue under the same later arrivals in two cases, with and without the admitted $y$ seconds of work.
  Under first-come-first-served order, a request arriving at time $t$ starts only after all unfinished work in the queue is done.
  The extra wait of this request is therefore the extra unfinished work $\Gamma(t)$ that the case with the admitted work holds at time $t$, which equals $y$ when the work is admitted.
  $\Gamma(t)$ changes over time, and a request arriving at a later time $t'$ waits $\Gamma(t')$ longer.
  With requests arriving at rate $\lambda_g$, the expected total extra wait of later requests is $\lambda_g\, \mathbb{E}[\int_0^\infty \Gamma(t)\, dt]$.
  Let $B$ be the first time at which the group becomes idle in the case without the admitted work.
  Since later requests keep arriving at load $\rho_g$ while the group serves the unfinished work, the busy time of the group extends by $\rho_g\, \mathbb{E}[B]$ on average.
  $B$ therefore satisfies $\mathbb{E}[B] = w + \rho_g\, \mathbb{E}[B]$, which gives $\mathbb{E}[B] = w / (1 - \rho_g)$.
  Hence, the time $B$ splits the integral into two parts,
  \begin{equation}
    \int_0^\infty \Gamma(t)\, dt = \int_0^{B} \Gamma(t)\, dt + \int_{B}^{\infty} \Gamma(t)\, dt .
    \label{eq:split}
  \end{equation}
  \textbf{1)~Existing work.}
  Before $B$, both cases serve the same work, giving $\Gamma(t) = y$ and thus
  \begin{equation}
    \mathbb{E}\Bigl[ \int_0^{B} \Gamma(t)\, dt \Bigr] = y\, \mathbb{E}[B] = \frac{y\, w}{1 - \rho_g} .
    \label{eq:part1}
  \end{equation}
  \textbf{2)~Admitted work.}
  After $B$, $\Gamma(t)$ decreases only while the case without the admitted work is idle, and later requests keep both cases busy for a fraction $\rho_g$ of the time.
  $\Gamma(t)$ therefore decreases from $y$ to zero at an average rate of $1 - \rho_g$, giving
  \begin{equation}
    \mathbb{E}\Bigl[ \int_{B}^{\infty} \Gamma(t)\, dt \Bigr] = \frac{y}{2} \cdot \frac{y}{1 - \rho_g} = \frac{y^2}{2\,(1 - \rho_g)} .
    \label{eq:part2}
  \end{equation}
  Adding the two parts gives
  \begin{equation}
    \lambda_g\, \mathbb{E}\Bigl[ \int_0^\infty \Gamma(t)\, dt \Bigr]
    = \frac{\lambda_g}{1 - \rho_g} \Bigl( y w + \frac{y^2}{2} \Bigr) = E_g(y) .
    \label{eq:total} \qedhere
  \end{equation}
\end{proof}

Since a satellite has $J$ groups of prefill nodes, we extend Proposition~\ref{prop:price} from one group to the $J$ groups of satellite $v$.
New requests join the group that frees first, and work added to a group therefore delays a later request only if the other $J - 1$ groups are busy, which happens with the Erlang C probability $P_{\mathrm{w}}(J{-}1, \rho_v)$, where $\rho_v$ is the mean load of the groups.
Each group receives a share $\Lambda_v / J$ of the request rate $\Lambda_v$ of the satellite, and the GPU price of $y$ seconds of work at satellite $v$ is $E_g(y)$ of~\eqref{eq:ext} with $\lambda_g = \Lambda_v / J$, $w = w_v$ and $\rho_g = \rho_v$, weighted by this probability,
\begin{equation}
\begin{aligned}
  E^{J}_v(y) &= P_{\mathrm{w}}(J{-}1, \rho_v)\, E_g(y) \\
  &= \frac{\Lambda_v}{J}\, P_{\mathrm{w}}(J{-}1, \rho_v)\, \frac{y\,(w_v + y/2)}{1 - \rho_v} .
\end{aligned}
  \label{eq:extj}
\end{equation}

\subsection{A Link Price for the Ring Lasers}
\label{sec:linkprice}

Similarly, the link price captures the extra wait that a fetch imposes on other KV cache transfers sharing the same lasers (Fig.~\ref{fig:problems}b).
Modeling each directed laser $e$ of rate $C_e$ as a first-come-first-served queue, a fetch of $S$ bytes adds $S / C_e$ seconds of work to laser $e$.
Laser $e$ is therefore a queue of the same form as a GPU group, and Proposition~\ref{prop:price} holds for laser $e$ with the work $y = S / C_e$.
Since the fetched KV cache crosses every laser on path $\pi$, we extend Proposition~\ref{prop:price} from one laser to path $\pi$ by summing the extra wait over these lasers,
\begin{equation}
  E^{\mathrm{link}}_\pi(S) = \sum_{e \in \pi} E_e\Bigl(\frac{S}{C_e}\Bigr) = \sum_{e \in \pi} \frac{\lambda_e\, (S/C_e)\, \bigl(w_e + S/(2C_e)\bigr)}{1 - \rho_e} ,
  \label{eq:linkprice}
\end{equation}
where $\lambda_e$ and $\rho_e$ are the transfer start rate and the utilization of laser $e$ in the latest load report.
The gateway estimates the queue $w_e$ of laser $e$ as it estimates the wait of a GPU group in~\eqref{eq:wait},
\begin{equation}
  w_e = \max\Bigl\{ w^{\mathrm{rep}}_e + (\rho_e - 1)\,\alpha_e ,\
  \frac{q_e}{2\,(1 - \rho_e)}, \ W^{u}_e / C_e \Bigr\} ,
  \label{eq:linkwait}
\end{equation}
where $w^{\mathrm{rep}}_e$ is the reported queue, $\alpha_e$ is the age of the report, $q_e$ is the rate at which transfers start on laser $e$ times the mean squared transfer time $(S/C_e)^2$ of a transfer, and $W^{u}_e$ is the bytes that the gateway itself has sent on laser $e$ that are not yet reported and still unsent.
To keep \eqref{eq:wait}, \eqref{eq:extj}, and \eqref{eq:linkprice} defined under overload, AstraLLM caps the load in each factor $1-\rho$ at $1-10^{-5}$.
\subsection{Spreading the Choices of Gateways}
\label{sec:spread}

Since all gateways share the same load reports, every gateway would decide the same satellite that has the lowest score defined in~\eqref{eq:decision} before the update from the next load report.
AstraLLM therefore randomly draws a satellite as the serving satellite from the distribution defined in~\eqref{eq:draw}, which prevents the satellite with the lowest score from being overloaded.

The draw follows from the errors that stale load reports leave in the scores.
While the wait $w_v$ in~\eqref{eq:wait} accounts for stale load reports, the projected unfinished work $w^{\mathrm{rep}}_g + (\rho_g - 1)\,\alpha_v$ adds only the expected work $\rho_g \alpha_v$ that arrives during the report's age $\alpha_v$.
The actual work that other gateways send varies around the mean and leaves random errors in the scores $s(v, x)$ of~\eqref{eq:decision}.
We model the score errors of the options as independent Gumbel variables with standard deviation $\theta$~\cite{gumbel1958}, whose scale parameter is $\beta = \frac{\sqrt{6}}{\pi}\,\theta$.
Under this model, the probability that satellite $v$ truly has the lowest score is proportional to $e^{-s(v, x)/\beta}$~\cite{mcfadden1974}, as shown in~\eqref{eq:draw}.
Thus, each satellite is weighted by $e^{-s(v, x)/\beta}$, where satellites whose scores differ by less than about $\beta$ receive comparable weights, spreading the requests of gateways to these satellites.

The standard deviation $\theta$ comes from two sources of staleness error, both of which grow with the age $\alpha_v$ of the load report.
\textbf{1)~Unseen work.}
Other gateways send work to satellite $v$ during $\alpha_v$, which is not included in the load report.
These arrivals form a compound Poisson process over $\alpha_v$ seconds, and counting each request at its mean work $\rho_g / \lambda_g$ gives the unseen work the variance $\sigma_w^2 = \alpha_v \rho_g^2 / \lambda_g$.
\textbf{2)~Stale load.}
The reported load $\rho_g$ averages the arrivals over the last $T_{\mathrm{win}}$ seconds.
By the time a turn arrives, this window has slid forward by $\alpha_v$ seconds, adding unseen arrivals and dropping the oldest $\alpha_v$ seconds of arrivals.
Because both $\alpha_v$-second intervals are independent compound Poisson sums, the reported satellite load $\rho_v$ differs from the current load with variance $\sigma_\rho^2$.
Both errors enter the GPU price in~\eqref{eq:extj} through $w_v$ and $\rho_v$.
Applying first-order error propagation (treating $P_{\mathrm{w}}$ as locally constant) yields
\begin{equation}
  \theta = \sqrt{
    \Bigl(\frac{\partial E_v^J}{\partial w_v}\Bigr)^{\!2} \sigma_w^2
    + \Bigl(\frac{\partial E_v^J}{\partial \rho_v}\Bigr)^{\!2} \sigma_\rho^2
  } ,
  \label{eq:theta}
\end{equation}
where
$\frac{\partial E_v^J}{\partial w_v} = \frac{\Lambda_v P_{\mathrm{w}}(J{-}1,\rho_v)\, y}{J\,(1-\rho_v)}$,
$\frac{\partial E_v^J}{\partial \rho_v} = \frac{w_v + y/2}{1-\rho_v}\frac{\partial E_v^J}{\partial w_v}$, and
\begin{equation}
  \sigma_w^2 = \alpha_v \frac{\rho_g^2}{\lambda_g} , \qquad
  \sigma_\rho^2 = \frac{2 \alpha_v}{(J\, T_{\mathrm{win}})^2}
  \sum_{g' \in v} \frac{\rho_{g'}^2}{\lambda_{g'}} ,
  \label{eq:sigmas}
\end{equation}
where $g$ is the minimizing group in~\eqref{eq:wait} and $g'$ ranges over all groups of satellite $v$.
AstraLLM evaluates $\theta$ at the option with the lowest score, so that all options of a turn share one scale $\beta$.

\noindent\textbf{Modeling approximations.}
To keep the scheduler closed-form and parameter-free, AstraLLM relies on two lightweight approximations:
(i) modeling prefill groups and ring lasers as independent FCFS queues and weighting the $J$-group GPU price by the Erlang~C probability $P_{\mathrm{w}}(J{-}1,\rho_v)$; and
(ii) modeling score errors as Gumbel-distributed and estimating $\theta$ via first-order error propagation on the GPU price terms in~\eqref{eq:theta} (treating $P_{\mathrm{w}}$ as locally constant, neglecting error covariance, and using single-group variance $\sigma_w^2$ for $w_v$).
Section~\ref{sec:eval} shows that these approximations are sufficient to keep TTFT close to the access-delay lower bound across loads.
\section{Evaluation}
\label{sec:eval}

We evaluate AstraLLM against six baselines across traffic loads, network conditions, and constellation sizes.  AstraLLM achieves the lowest mean TTFT up to saturation, sustaining 1.7$\times$ the request rate of SGLang and 2.3$\times$ that of Mooncake (Section~\ref{sec:main}).  We also break down each policy's TTFT (Section~\ref{sec:breakdown}), test the sensitivity to laser bandwidth and report intervals (Section~\ref{sec:sensitivity}), ablate each design component (Section~\ref{sec:ablation}), and evaluate scaling across fleet sizes (Section~\ref{sec:fleet}).
\begin{figure*}[t]
\centering
\includegraphics[width=\textwidth]{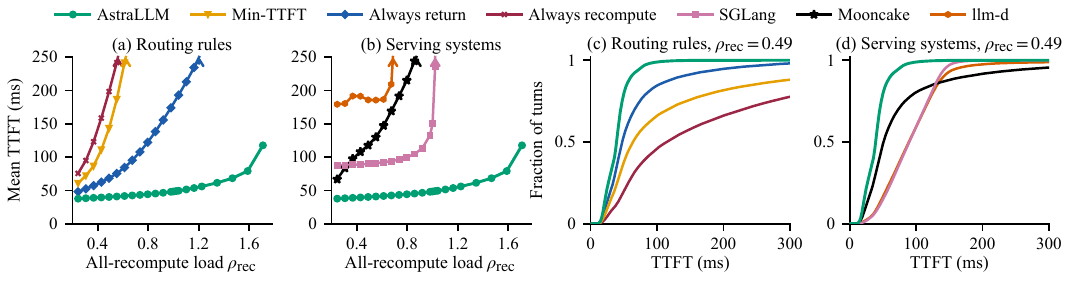}
\caption{AstraLLM achieves the lowest mean TTFT among all seven policies up to saturation.  (a,~b)~Mean TTFT versus normalized all-recompute load; (c,~d)~TTFT CDFs at $\nrate = 0.49$, the highest load at which all seven policies remain stable.}
\label{fig:load}
\end{figure*}
\subsection{Setup}
\label{sec:setup}
\begin{table}[t]
\caption{Parameters of the modeled constellation}
\label{tab:params}
\centering
\footnotesize
\begin{tabular*}{\columnwidth}{@{}l@{\extracolsep{\fill}}cr@{}}
\toprule
Quantity & Symbol & Value \\
\midrule
Satellites per ring & & 288 \\
Ring altitudes~\cite{spacex2026grant} & & 570 and 580 km \\
Ring-hop propagation delay & $\tau_h$ & 0.505 ms \\
Rate of a ring laser~\cite{starlink2026v3} & $C_e$ & 400 Gb/s \\
GPUs per satellite~\cite{starmind2026nvidia,nvidia2026nvl72} & & 72 \\
Default prefill groups per satellite & $J$ & 4 \\
GPUs per prefill group~\cite{kimi2025k2,nvidia2026nvl72} & & 8 \\
Prefill time per token~\cite{zhong2024distserve,kimi2025k2,deepseek2025infer,nvidia2026nvl72} & $b$ & 3.12 \textmu{}s \\
Prefill time per token pair~\cite{zhong2024distserve,kimi2025k2,deepseek2025infer,nvidia2026nvl72} & $c$ & 0.124 ns \\
KV cache per token~\cite{kimi2025k2} & & 70 KB \\
Decode GPUs per satellite & & 40 \\
Routed experts per decode GPU/layer~\cite{kimi2025k2} & & 10 \\
HBM per GPU~\cite{nvidia2026nvl72} & & 288 GB \\
HBM bandwidth per GPU~\cite{nvidia2026nvl72} & & 19.2 TB/s \\
NVMe storage per satellite~\cite{lenovo2026nvl72,micron9650} & & 553 TB \\
Drive read rate of a group~\cite{lenovo2026nvl72,micron9650} & $R_{\mathrm{ssd}}$ & 224 GB/s \\
Report period & $T_{\mathrm{rep}}$ & 0.1 s \\
Window of reported loads & $T_{\mathrm{win}}$ & 60 s \\
Median shared system prompt~\cite{wang2025kvwild} & $n_0$ & 64 tokens \\
Median think time~\cite{wang2025kvwild} & & 106 s \\
Start of the simulated hours & & 00:00, 12:00 UTC \\
\bottomrule
\end{tabular*}
\end{table}

\textbf{Constellation and demand.}  We evaluate AstraLLM with a custom packet-level simulator built in ns-3~\cite{riley2010ns3}.  Beyond the relay shell of Hypatia~\cite{kassing2020hypatia}, we add the compute rings, the links between the two layers, and an LLM serving model covering every prefill step, decode step, KV cache transfer, and load report.  Requests and KV cache transfers travel as packets that share the link queues of the moving constellation.  Existing simulators do not cover this setting, as they either model satellite networks without LLM serving~\cite{kassing2020hypatia} or target single terrestrial clusters.  We instantiate the Starlink relay satellites from CelesTrak orbital element sets~\cite{kelso2020gp}, propagate them with SGP4, and connect them using Hypatia's +Grid topology~\cite{kassing2020hypatia}.  The simulator updates ISL connectivity every 15~s to match Starlink's reconfiguration interval~\cite{wu2025sate} and computes shortest paths between satellite pairs using Dijkstra's algorithm.  KV cache transfers use a receiver-driven transport modeled on Homa~\cite{montazeri2018homa}.  User demand enters the network through 260 operational Starlink gateways~\cite{starlinkinsider2026gateways} with visibility to the shell, with traffic from cities of at least 15,000 residents~\cite{geonames} routed to the nearest gateway.  Each city's demand follows its local time-of-day pattern and is scaled by country-level generative AI adoption from Microsoft's report~\cite{msft2025aidiffusion}.  We simulate two start times (Table~\ref{tab:params}) so that the evening traffic peaks of different continents fall beneath the compute rings.

\textbf{Workload.}  Multi-turn sessions are drawn from a two-hour segment of the Alibaba Qwen Bailian production chat trace~\cite{wang2025kvwild}.  New sessions arrive as a Poisson process, making the normalized load $\nrate$ proportional to the overall request rate.  Each subsequent turn arrives after the previous response finishes decoding, plus the think time from the trace.  Each run uses a 10-minute warm-up followed by a 15-minute measurement window.

\textbf{Model and hardware.}  Each compute satellite runs Kimi K2~\cite{kimi2025k2} on a single Vera Rubin NVL72 rack~\cite{nvidia2026nvl72}.  Prefill execution time~\eqref{eq:tpf} follows the DistServe performance model~\cite{zhong2024distserve}, where constants $b$ and $c$ are derived from Kimi K2's per-token and per-token-pair FLOPs divided by the FP8 throughput of a prefill group.  We assume the same hardware utilization as DeepSeek-V3's production prefill deployment~\cite{deepseek2025infer}, which shares Kimi K2's architecture.

\textbf{Decode.}  The 40 decode GPUs on each satellite serve Kimi K2 using
data-parallel attention and expert parallelism, matching DeepSeek-V3's decode
setup~\cite{deepseek2025infer}, and each decode step lasts as long as reading
the model weights and the KV caches of its batch from HBM~\cite{nvidia2026nvl72}.
Each satellite lends a prefill group to its decode instance when decode steps
exceed a latency target or answers wait for KV cache room and the remaining
groups can absorb the prefill load, and takes it back when prefill queues build
up or decode no longer needs it, resizing the instance as in vLLM's elastic
expert parallelism~\cite{vllm2026elasticep}.

\textbf{Baselines.}  We compare AstraLLM against three heuristic policies and three adapted terrestrial serving systems.  The three heuristics choose among AstraLLM's actions using the same load reports:
\emph{Min-TTFT} selects the action with the lowest estimated TTFT;
\emph{Always Return} routes every resumed turn back to its cache holder; and
\emph{Always Recompute} routes every turn to the satellite with the lowest sum of access round-trip time and GPU queuing delay.

To represent how terrestrial serving systems manage KV caches, we also deploy three production designs in orbit, treating each satellite as a worker and placing one router at each gateway. We port each system by keeping its decision rule unchanged and giving its router at each gateway the same periodic load reports as AstraLLM. 
\emph{SGLang}~\cite{zheng2024sglang} never moves a KV cache.  Each gateway records in one prefix tree which satellite it sent each prompt to, and sends a turn back to the satellite holding at least 30\% of its prompt.  Otherwise, or when its own requests pile up unevenly, it picks the satellite with the fewest of them.  Because all sessions share the tree, turns with the same system prompt crowd onto one satellite until that rebalancing moves them.
\emph{Mooncake}~\cite[Alg. 1]{qin2025mooncake} sends each turn to the satellite with the lowest estimated TTFT.  For a later turn, the satellite holding the session's cache reuses it and every other satellite fetches it from that holder, so Mooncake does not recompute a cached turn.
\emph{llm-d}~\cite{llmd2026p2p} assigns each turn to the worker with the fewest remaining prefill tokens and pulls the KV cache from the holder whenever the holder's cached prefix is at least 2,048 tokens longer.

\subsection{End-to-End TTFT and Capacity}
\label{sec:main}

\begin{figure}[t]
\centering
\includegraphics[width=\columnwidth]{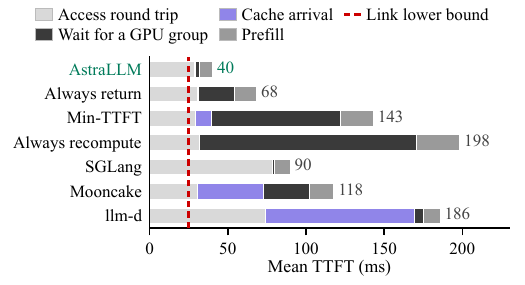}
\caption{Mean TTFT breakdown into four components at $\nrate = 0.49$.  Only AstraLLM keeps both queuing delays and access round-trip time low.}
\label{fig:breakdown}
\end{figure}

\begin{figure*}[t]
\centering
\setlength{\abovecaptionskip}{1pt}
\begin{minipage}[t]{0.475\textwidth}
\centering
\includegraphics[width=\linewidth]{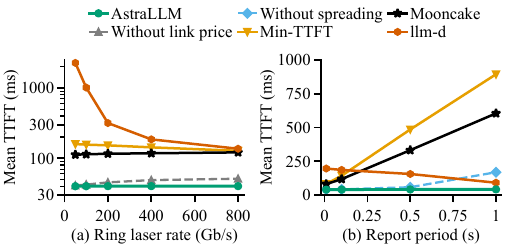}
\caption{Mean TTFT at $\nrate = 0.49$ versus (a)~ring laser bandwidth and (b)~load report period.}
\label{fig:sensitivity}
\end{minipage}\hfill
\begin{minipage}[t]{0.29\textwidth}
\centering
\includegraphics[width=\linewidth]{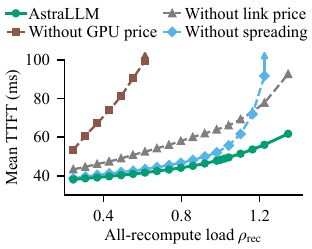}
\caption{Ablation study.}
\label{fig:ablation}
\end{minipage}\hfill
\begin{minipage}[t]{0.215\textwidth}
\centering
\includegraphics[width=\linewidth]{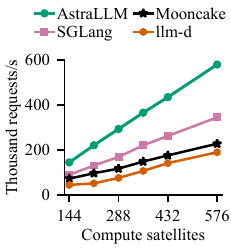}
\caption{System capacity.}
\label{fig:fleet}
\end{minipage}
\end{figure*}

AstraLLM achieves the lowest mean TTFT across all loads up to its saturation point at $\nrate = 1.71$ (Fig.~\ref{fig:load}a,~b).  At $\nrate = 0.49$, the highest load at which all seven policies remain stable, AstraLLM's entire TTFT distribution lies strictly to the left of every baseline (Fig.~\ref{fig:load}c,~d).

Four baselines (Always Return, Mooncake, Min-TTFT, and Always Recompute) saturate early because they ignore the queuing delay a turn imposes on subsequent arrivals.  Always Recompute and Min-TTFT overload nearby prefill groups with redundant computation, while Always Return pins high-demand regions to the same holders.  By combining cache reuse with congestion-aware offloading, AstraLLM sustains 1.4$\times$ to 3.5$\times$ their maximum stable request rates (2.3$\times$ over Mooncake and 3.1$\times$ over Min-TTFT), while reducing mean TTFT by 2.9$\times$ over Mooncake and 2.8$\times$ over Min-TTFT across their stable ranges.

SGLang and llm-d avoid localized GPU hotspots by spreading traffic across the entire ring, but this global spread incurs high transfer latency and still hits secondary bottlenecks.  SGLang's load-balancing rule redirects turns away from their holders, forfeiting cache reuse and eventually saturating its prefill groups; AstraLLM sustains 1.7$\times$ SGLang's maximum stable rate and reduces mean TTFT by 2.5$\times$.  Conversely, llm-d avoids GPU saturation by pulling caches to idle satellites, but its unpriced fetches build unbounded queues on the ring lasers at moderate load, resulting in 4.7$\times$ higher mean TTFT than AstraLLM.

\subsection{TTFT Breakdown}
\label{sec:breakdown}

Fig.~\ref{fig:breakdown} decomposes mean TTFT into four components at $\nrate = 0.49$.  AstraLLM's 40~ms mean TTFT consists mostly of the unavoidable 29~ms access round-trip delay, with minimal queuing or transfer overhead.  Interestingly, Always Return has longer prefill time than AstraLLM, even though it recomputes the fewest tokens.  This happens because pinning sessions crowds hotspot satellites, which increases batch sizes and slows down each prefill step.  The same effect appears across Always Return, Min-TTFT, and Always Recompute, whose prefill times grow progressively longer as GPU load rises.  Although Min-TTFT achieves an access round-trip delay similar to AstraLLM, it still suffers from severe GPU queuing and cache-arrival delays, showing why congestion-aware scheduling is necessary.

In contrast to these local baselines, SGLang and llm-d incur much higher access round-trip delays.  Both systems apply terrestrial load balancing across the entire ring without accounting for satellite propagation time, trading lower GPU queuing for much longer network paths.  Mooncake and llm-d also reflect terrestrial assumptions by pulling KV caches from remote servers.  Under limited ISL capacity, these unpriced fetches place heavy pressure on the ring lasers and result in large transfer delays.

\subsection{Sensitivity to Link Bandwidth and Report Period}
\label{sec:sensitivity}

AstraLLM achieves the lowest mean TTFT across all tested ring laser rates and report periods, staying nearly flat at 40~ms from 50 to 800~Gb/s (Fig.~\ref{fig:sensitivity}a).  Surprisingly, higher laser bandwidth actually raises mean TTFT for Mooncake and the variant without link pricing.  Without link pricing, faster lasers make remote fetches look cheaper, triggering more frequent and larger transfers that increase queuing delay on the ring lasers.  By pricing concurrent transfers, AstraLLM avoids this over-fetching at every bandwidth setting.

As the report period increases from 0.01~s to 1~s, stale reports sharply degrade deterministic policies such as Mooncake and Min-TTFT, pushing Mooncake's mean TTFT from 81~ms to 605~ms (Fig.~\ref{fig:sensitivity}b).  By contrast, llm-d's mean TTFT decreases under longer periods because its official router resets its round-robin tie-breaking counter whenever the eligible worker set changes, causing frequent reports to synchronize gateways onto the same satellites repeatedly.  Over the same range, AstraLLM's mean TTFT rises only from 40~ms to 42~ms, as its randomized selection spreads traffic more broadly when reports become older.  Without randomized spreading, AstraLLM's mean TTFT instead climbs to 169~ms at 1~s, as gateways that read the same stale report crowd onto the same satellites.

\subsection{Ablation Study}
\label{sec:ablation}

Fig.~\ref{fig:ablation} evaluates the contribution of each component in AstraLLM by removing them one at a time.  GPU pricing provides the largest benefit.  Without it, turns overload busy prefill groups and cause mean TTFT to rise rapidly, reaching 1.9$\times$ that of full AstraLLM even at $\nrate = 0.49$.  Link pricing improves TTFT across all loads, though its impact is smaller than that of GPU pricing.  Removing it adds a 15\% penalty even at the lowest load as unpriced cache transfers contend on the ring lasers, and the gap widens as load increases.  Finally, randomized spreading delivers its primary benefit at high load.  At low load, satellites have enough headroom to absorb requests even when multiple gateways select the same target, so removing spreading has little effect.  As load rises, however, this crowding intensifies and causes TTFT to surge, increasing mean TTFT by 277\% at $\nrate = 1.35$.

\subsection{System Capacity}
\label{sec:fleet}

A request rate is stable if, on every satellite, the load of the busiest prefill group and the decode load both stay below 1, where the decode load is the arrival rate of decode tokens divided by the maximum throughput of the decode GPUs, including the prefill groups the satellite can lend. AstraLLM achieves the highest stable request rate across all tested constellation sizes, and its lead over every baseline widens as more compute satellites are added (Fig.~\ref{fig:fleet}).  In our default constellation, AstraLLM serves 1.7$\times$ and 2.3$\times$ more requests per second than SGLang and Mooncake, respectively.  In orbital data centers, where launch and space-qualified hardware dominate deployment cost, this capacity gap translates directly into major infrastructure savings.  Matching the throughput that AstraLLM delivers with 288 satellites requires more than 432 satellites under SGLang, and even more under the other baselines.  Effective scheduling thus cuts the required constellation size by at least a third, avoiding the cost of building, launching, and operating over a hundred additional compute satellites.
\section{Related Work}
\label{sec:related}

\noindent\textbf{KV-cache reuse in LLM serving.}
Production LLM traffic is mostly multi-turn conversation, where new prompts repeat previous turns, allowing over half of all input tokens to reuse an existing KV cache~\cite{wang2025kvwild}.
Disaggregated serving moves the cache from prefill to decode instances~\cite{zhong2024distserve}, and single-server systems keep, prefetch or partly recompute the cache of an idle session~\cite{gao2024cachedattention,jin2025cake}.
Pallas combines KV transfer with recomputation when a mobile user hands over between cells~\cite{ding2026pallas}.
In all of these systems the servers are static, whereas the servers of an orbital data center move, adding an extra transfer cost to the reuse of the KV cache.

\noindent\textbf{Request scheduling.}
LLM serving systems schedule requests according to the token length of each request and the utilization of each GPU.
SGLang routes each request to the instance with the longest cached prefix unless the instances become unbalanced~\cite{zheng2024sglang}.
Mooncake picks the instance of least estimated TTFT, with or without a KV transfer~\cite{qin2025mooncake}.
These schedulers score each request by its own cost, either the cache it hits or the TTFT it sees, and treat every instance as equally far from the user.
In an orbital data center, however, the distance to each satellite changes between turns, and none of these schedulers accounts for the delay a recomputation imposes on later requests.

\noindent\textbf{LLM inference in orbit.}
Recent works have started to study LLM inference on satellites.
Clockwork provisions stateless workflows~\cite{gravara2026clockwork}, and SpaceMoE splits the layers of a model across satellites~\cite{wang2026spacemoe}.
Space AI Infrastructure keeps each request on one compute cluster and re-routes only the token stream at a handover~\cite{wang2026spaceai}, and OrbitWeaver decides between storing and recomputing the cache after an interruption~\cite{huang2026orbitweaver}.
None of these works serves multi-turn conversations whose cache holder drifts away from the user between turns, and AstraLLM is the first scheduler to choose the serving satellite and the KV cache action for each turn.

\section{Conclusion}
\label{sec:conclusion}

In orbit, the holder of a session's KV cache drifts away while the user thinks, adding only a few milliseconds to a single turn.  Instead, the drift turns scheduling into a load problem, where recomputing overloads GPU groups, fetching caches congests ring lasers, and stale reports cause gateways to crowd the same satellite.  AstraLLM charges each turn for the delay its GPU work and cache transfers impose on later requests, and spreads gateway choices in proportion to the uncertainty of those estimates.  Removing any component reintroduces the problem it addresses.  Together, these mechanisms reduce mean TTFT by 2.5$\times$ compared with the strongest serving system and serve 1.7$\times$ and 2.3$\times$ the request rate of SGLang and Mooncake, respectively.  Because every compute satellite carries steep launch and hardware costs, this capacity gain lets operators serve the same global demand with at least a third fewer satellites in orbit.

{\footnotesize
\bibliographystyle{IEEEtran}
\bibliography{refs_main}}
\end{document}